\documentclass[a4paper]{article}
\usepackage[utf8]{inputenc}
\usepackage{subcaption}
\usepackage{graphicx}
\usepackage{authblk}
\usepackage{amsmath}
\usepackage{amssymb}
\usepackage{xcolor}
\usepackage{biblatex} 
\usepackage{hyperref}
\usepackage{amsthm} 

\newtheorem{theorem}{Theorem}[section]

\theoremstyle{definition}
\newtheorem{corollary}[theorem]{Corollary} 
\newtheorem{lemma}[theorem]{Lemma} 

\theoremstyle{definition}
\newtheorem{definition}[theorem]{Definition}

\theoremstyle{definition}

\theoremstyle{definition}

\theoremstyle{definition}

\theoremstyle{remark}
\newtheorem*{remark}{Remark}

\theoremstyle{remark}

\title{Chiral measurements}
\author[1]{Christopher~G.~Lester }
\affil[1]{
Cavendish Laboratory,
University of Cambridge,
CB3 0HE,
United Kingdom
}
\date{1st November 2021}

\begin{document}

\maketitle

\begin{abstract}
Searches for parity violation at particle physics collider experiments without polarised initial states or final-state polarimeters lack a formal framework within which some of their methods and results can be efficiently described.  This document defines  nomenclature which is intended to support future works in this area, however it has equal relevance to searches concerned with more conventional (and already well supported) pre-existing tests of parity-violation since it can  be viewed as a formal axiomatisation of process of gaining evidence about parity violation regardless of its cause.
\end{abstract}

\section*{Foreword}

\scriptsize 


Although the results presented in this  work were all independently derived 
for use in particle physics,
it  became clear in the process of writing-up that many of its key theorems and concepts (or very closely related ones) have already been published a number of years ago in other fields -- mostly in disciplines relating to molecular chirality and chemistry.\footnote{For example: the second half of Section 2.1 of \cite{PetitjeanReview} mentions results that overlap with many  herein; Figure 6 of \cite{RevModPhys.71.1745} performs largely the same function as Figures~\ref{fig:handofdeath} or \ref{fig:blobofdeath} herein; there is much discussion of `chiral zeros' in \cite{mislowweinberg} which relates to discussions surrounding `non ideal parities' described herein; and other forms of overlap can be found withing parts of \cite{doi:10.1080/00268970902835611,ruch1970}.  Quite possibly there is even overlap with remarks Lord Kelvin may have made on Chirality when delivering hes Baltimore Lectures in 1884 \cite{KelvinBaltimore}.}  The terminology used in those other papers is not always the same as here, and nor is the context identical. Nonetheless, there is usually some form of equivalence that can translate a result from one field into the other.  The most direct example of this is that the three-momenta of particles in LHC events has a largely similar role to that played by the position-vectors of parts of molecules in the chemistry literature.
For every rule there is an exception, however, and the author has not found an equivalent of Theorem~\ref{thm:onlyneedoddchiralfuncs} or its Corollary~\ref{cor:ffhkdfffyeijwhdf} elsewhere.  Consequently some results may indeed be novel.

The high degree of overlap between the pre-existing chemical literature on chirality and my work in particle physics had led me to believe that there was little point in continuing with the present work as a publication.  However, feedback from colleagues who heard that this work was to due be `shelved' has encouraged me to reconsider. They have argued that there is presently a dearth of LHC literature related to tests of parity violation which can be performed at the LHC\footnote{At present there are only two known to the author: \cite{Lester:2019bso} and \cite{Lester:2020jrg}.}, and that the chances of that area growing are low if there do not exist papers which set out the necessary frameworks and terminology in forms which are accessible to  the intended audience. This work is therefore made public  in the hope it will assist future works in collider physics, rather than on account of any significant novelty in the strictest sense.

Despite the intended link to particle physics, very few `important' references to particle physics will actually be found in this paper. This strange state of affairs derives from a desire to keep the formulation herein sufficiently general that it can be used in a large number of future particle physics contexts  -- and those contexts are sufficiently diverse that the the definitions used herein have been kept as abstract as possible.  The `big' links to particle physics will only be apparent in the future works which intend to use the nomenclature and theorems defined herein.

\normalsize 

\section{Introduction}
Some objects are distinguishable from their reflections (or transformations under parity) and some are not.  Those which are are termed `\textit{chiral}', and the others `\textit{achiral}'. A person's right hand is chiral as it cannot be brought into co-incidence with its reflection by mere rotations or translations, while a person's nose is typically  achiral  because it looks the same as its mirror image. Many things which appear achiral at one scale may turn out to be chiral when inspected at another.  For example, a small pimple above one nostril could make a nose asymmetrical that would otherwise have appeared symmetrical from afar.  Chirality-vs-achirality  therefore need not be a simple black-and-white distinction but is something which can have shades of grey.  It is something whose size may need to be measured in different ways at different scales, and for which the degree of chirality may depend considerably on the method of measurement used.  

Within particle physics, measurements concerning parity and chirality  sometimes rely on easily and directly observable characteristics of particles emerging from collisions, such as their momenta, energies, masses and positions.  Measurements can also rely on external quantities (such the directions of electric or magnetic fields, or a beam axis) or in collision inputs (beam energies, directions,  masses and other properties of the colliding particles, etc.) or on properties inferred from pieces of apparatus (e.g.~polarisers) or from daughter products of parent particles.

The measurements which discovered parity violation \cite{PhysRev.105.1413} relied on at their core on two inputs: (i) the directions, $\hat {\vec n}$, of electrons emitted in beta decay from a cobalt-60 source, and (ii) the direction, $\vec B$, of a magnetic field that had been used to align the spins of the cobalt-60 nuclei.  In a gross oversimplification of the work of Wu and others in that study, the expected value of the dot product of those quantities, $ \hat {\vec n} \cdot \vec B$ (with the expectation taken over all observations)  was predicted to be zero in the absence of parity  violation. Observations, however, showed the dot product was usually negative. The rest is history.

One way of looking at this historical measurement is to view the experimenters as having recorded a series of events $x_1, x_2,\ldots\in X $, where each event $x$ consisted of an observed electron direction and the associated magnetic field: $x=(\hat {\vec n},\vec B)$. One could then consider the experimenters to have created a measuring function $m(x)$ which returns the dot product of the two vectors contained within each $x$, i.e.:
\begin{align}
m(x)
&=
m((\hat {\vec n},\vec B))
\equiv
\hat {\vec n} \cdot \vec B.
\label{eq:wuobs}
\end{align}%
Finally we can imagine the experimenters having plotted a histogram of values $m(x_1), m(x_2), ....$ for each observation in their dataset, and having noticed that the histogram was asymmetric (i.e.~with more negative than positive values).  The reason such an asymmetry was interesting to them (and is to us) is that their measurement function $m$  was parity-odd (meaning that the signs of its outputs would have been inverted if  observations $x$ sent to it had been viewed in a mirror as they were recorded)\footnote{The action of the parity transformation would be to replace $\hat {\vec n}\rightarrow -\hat {\vec n}$ since it is a direction.  The parity transformation would leave the magnetic field alone, however, $\vec B \rightarrow \vec B$, since the currents in the solenoids generating it would (after a transformations $x\rightarrow -x$,  $y\rightarrow -y$, and $z\rightarrow -z$) still  circulate the same way.  $\vec B$ is therefore said to be a `pseudovector', while $x$ is said to be a `vector'.
} and so if the laws of physics were parity-invariant then there should have been equal probability for generating positive values of $m$ as negative values of $m$.

The example just given is an example of a parity-odd measurement being used to tell us something about parity. But parity-even measurements are also used from time to time.  The forward-backward asymmetry of the $Z$-boson at the Large Electron Positron collider (LEP) is one such, and is often cited as a source of the evidence that the $Z$-boson couples differently to the left and right chiral components of electrons or muons.\footnote{The whole of this discussion only concerns $e^+e^-\rightarrow \mu^+\mu^-$ in the context of collisions at LEP with unpolarised beams.  Some parts of the discussion would become false if the beams were to acquire measurable polarisations.} In this case: (i) the observations $x$ in events of the form $e^+e^-\rightarrow Z \rightarrow \mu^+\mu^-$ can each be considered to be an ordered list of the four-momenta of each of the particles (measured in a consistent frame), i.e.~$x=(
p_{e^-}^\mu, 
p_{e^+}^\mu ,
p_{\mu^-}^\mu, 
p_{\mu^+}^\mu
)$, or $x=(a^\mu,b^\mu,c^\nu,d^\nu)$ for short; and (ii)  the measurement function returns the cosine of the angle between the incoming electron $a^\mu$ and the outgoing muon $c^\mu$ in the lab frame, i.e.~the frame in which $\Lambda^\mu=a^\mu+b^\mu$ is at rest.  This measurement function may be written in terms of four momenta as follows:
\begin{align}
m(x) &= \frac
{(a\cdot\Lambda)( c\cdot \Lambda) - (a\cdot c)\Lambda^2}
{
\sqrt{
(a\cdot\Lambda)^2-a^2 \Lambda^2
}
\sqrt{
(c\cdot\Lambda)^2-c^2 \Lambda^2
}
}\label{eq:notwuobs} 
\end{align}%
where $\Lambda^\mu = a^\mu+b^\mu$.
Whereas the measurement function $m$ of Eq.~\eqref{eq:wuobs} was parity-odd, the $m$ of Eq.~\eqref{eq:notwuobs} is plainly parity-even since it is composed entirely of Lorentz products between four-vectors (not pseudo four-vectors) and so it does not change sign under parity. Nevertheless, the distribution of $m(x)$ over events still captures interesting information about parity, not least because the forward-backward asymmetry  $A_{\mathrm{FB}}$ defined by \begin{align}
    A_{\mathrm{FB}} &= \frac{
    \sigma_{m(x)>0} - \sigma_{m(x)<0}
    }{
        \sigma_{m(x)>0} + \sigma_{m(x)<0}
    }
    \label{eq:AFB}
\end{align}
is equal (at least at first order, and with observations in a $4\pi$ detector) to the quantity:
$$
A=\frac 3 4 A_e A_\mu$$ where $$
A_f=\frac
{(c_L^f)^2 - (c_R^f)^2}
{(c_L^f)^2 + (c_R^f)^2}
$$ and 
$c_L^f$ and $c_R^f$ are (respectively) the left and right chiral couplings of the $f$-lepton to the $Z$-boson.  It may thus be seen that a non-zero value for the forward-backward asymmetry $A_{\mathrm{FB}}$  which is observed in data provides evidence that the left and right muon (and also the left and right electron) do not couple equally to the $Z$-boson.

With usage of parity-even and parity-odd measurements abounding in particle physics, it is natural to ask why some measurements are of one type and some of the other, and whether one type of measurement contains any information that is not present in the other (or vice versa). 

This paper seeks to lay down an axiomatic framework so that  questions similar to those above can be asked very precisely in different contexts. A benefit of having such a framework is that theorems can be proved which can allow strong statements to be made in certain circumstances.    Indeed: one important theorem we prove herein tells us that in a wide set of circumstances (which are not very onerous, and are likely to be satisfied in any particle-physics collider scenario) there is, in principle, no information contained within parity-even measurements that is not contained within parity-odd measurements.\footnote{Indeed, the extra information in the parity-odd measurements of \cite{PhysRev.105.1413} are precisely why they could claim discovery of parity violation and get a Nobel Prize for it, while measurements at LEP of the forward-backward asymmetry of the $Z$-boson with unpolarised beams would not have bean able to lead to the same outcome if they had come first. The reason is that the observed asymmetry in $A_{\textrm{FB}}$, although explained by the parity-violating $c_L$ and $c_R$ coefficients of the weak interaction in the Standard Model, could also (in principle) be explained by a different  parity `respecting' theory.  For example, if it were simply the case that `other things being equal, particles in the final state of a $2\rightarrow 2$ scattering process  tend to prefer to go the same way as the incoming particles of the same charge as themselves', then an asymmetry in $A_{\textbf{FB}}$ could easily be generated. Therefore the vanilla forward-backward asymmetry measurement, although often cast as providing evidence for parity violation, really does no such thing. It \textit{is} able to constrain parity-violating parameters within a theory which is already assumed to represent reality and which is already assumed to violate parity. But it does not, on its own, compel us to believe that the laws of physics violate parity in the same way that the Wu measurements do.}

The above result may seem obvious and not really in need of a long paper.  After all: it is plain that discarding the signs of a parity-odd measurement will give us a parity-even measurement, and so it does not seem unreasonable to image  that every parity even measurement could be construed as some sort of  parity-odd measurement whose sign was discarded somehow.  But even though that result `feels' sensible, is is already apparent from the example $m(x)$ given in Eq.~\eqref{eq:notwuobs} that the process of `discarding' signs cannot be as simple as taking a modulus since the quantity $m(x)$ in Eq.~\eqref{eq:notwuobs} can take both positive and negative values, and both of those values are very important for defining the asymmetry $A_{\textrm{FB}}$
in Eq.~\ref{eq:AFB}. And this is not the only difficulty.  There are important questions about what how to define achirality when there are other symmetries (e.g. translations and rotations) which may or may not matter for the purposes of deciding whether an object is or is not distinguishable from its mirror image.

This work exists to provide a framework for precisely addressing issues like the above.   The amount of truly original material contained within is small.  But the intention is that the framework will show dividends in later works which will begin with \cite{paper2} and \cite{paper3}.

\subsection{Structure of the rest of the paper}

Section~\ref{sec:deftermtheor} contains the majority of the general definitions, terminology and theorems related to chiral. It concludes with the statement and proof of Theorem~\ref{thm:onlyneedoddchiralfuncs} and a discussion of the messages we should take home from its Corollary~\ref{cor:ffhkdfffyeijwhdf}.

Section~\ref{sec:nowweareonapritys} then focuses on parity-odd chiral measures entirely (even giving them the special name of `parities') motivated by the proof in the previous section of the circumstances in which doing so can be done without any loss of generality.    Much of the discussion in Section~\ref{sec:nowweareonapritys} is (deliberately) a little more discursive and less rigorous than in the preceding section, since it explores notions related to continuity which, if discussed rigorously, would vastly increase the length of the paper.  Continuity aside, the most important concept introduced in Section~\ref{sec:nowweareonapritys} is that of an `ideal parities'.

Section~\ref{sec:exampleseectreewre} discusses some concrete examples which  illustrate some of the concepts introduced in the paper.  The longest of these (involving star-domains) is constructed entirely for its pedagogical value.  Another example describes the connections between actual work in particle physics and some of the concepts introduced in this paper.

\section{Definitions,   Terminology and Theorems}
\label{sec:deftermtheor}

\subsection{Measurements}

\begin{definition}[Measurements]
We \label{def:measurement} refer to a function $m$ as a `\textit{measurement}' if and only if it maps any object $x\in X$  to a value $m(x)$ in a vector space $V$.  
\begin{remark}
A measurement $m$ so defined need have no explicit connection to parity. For example: if $x=(E,p_x,p_y,p_z)$ were the reconstructed four-momentum of a particle  seen in a collider physics experiment, then $m$ could in principle be a function which returns a vector  $v\in V=\mathbb{R}^3$ containing: (i) the mass of the $x$-particle, (ii) its energy added to thrice its $z$-momentum, and (iii) the number 8: $$m(x)=\left(\sqrt{E^2-p_x^2-p_y^2-p_z^2},\ E+3 p_z 
,\ 8\right).$$ 
\end{remark}
\begin{remark}
$V$ has been called a vector space (rather than just a space of ordered tuples) because we later multiply measurements by dimensionless numbers, negate measurements, add measurements, subtract measurements, etc.  This requirement is  not very `restrictive' as any ordered tuples of numbers can support the above operations so long as each element of the tuple retains a consistent dimension (mass / length / speed, etc.) for any $x$.  For example, the vector space requirement would ban a measurement $m$ from  assigning to its first component the quantity `$p_x$ when $p_x>0$ but -3 otherwise', since one could not then add $m(p_x=+4~\mathrm{GeV})$ to $m(p_x=-5~\mathrm{GeV})$ in a dimensionfully valid way.  Nothing, however, would prevent $m(p_x)$ being defined to be $(p_x,0)$ if $p_x>0$ and $(0, -3)$ otherwise, since the first component is always a momentum and the second component is always dimensionless nomatter what the input.   
\end{remark}
\end{definition}
\begin{definition}[Equivalent measurements]
Informally: two \textit{measurement}s $m_1$ and $m_2$ are termed \textit{equivalent} if the results of either one can be deduced from the results of the other.  More formally: if $m_1:X\rightarrow V_1$ and $m_2:X\rightarrow V_2$ then $m_1$ and $m_2$ are termed \textit{equivalent} if there exists a function $f_{12}:V_1\rightarrow V_2$ and a function $f_{21} :V_2\rightarrow V_1$ such that $m_2(x) = f_{12}(m_1(x))$ and $m_1(x)=f_{21}(m_2(x))$ for all $x\in X$. \label{eq:equivalentmeasurements}
\begin{remark}
By the above definition, measurements made in the Kelvin, Fahrenheit and Celsius temperature scales would be  equivalent as a measurement made in one can be converted into a measurement made in any other.  The transformations which interrelate those temperature scales happen to be linear but the definition of equivalence does not require this.
\end{remark}
\begin{remark}
Definition~\ref{eq:equivalentmeasurements} is not equivalent to requiring that there be a bijection between $V_1$ and $V_2$ since it may be the case that the image of $m_i$ occupies only part of its codomain $V_i$.  It does, however, mean that there is a bijection between the images of $x$ in $V_1$ and $V_2$.
\end{remark}
\begin{definition}[$|m|$: the absolute form of a measurement $m$]
If $m$ is a \textit{measurement} in a vector space over real and/or complex fields  then $|m|$ denotes the (new) measurement obtained from $m$ by taking the modulus \label{def:absoluteformof} of each component of $m$.  For example: if $x\in \mathbb{R}$ and $m(x)=(\sin x, x^3, e^{i x})\in\mathbb{R}^2\otimes\mathbb{C}$ then $|m|(x) =(\left|{\sin x}\right|,  |x|^3,1)$. The measurement $|m|$ is termed the \textit{absolute form of the measurement $m$}.
\end{definition}
\subsection{Parity}
\end{definition}
\begin{definition}[Parity operator]
We define $P$ to be a \textit{parity operator} if it maps any $x\in X$ to another point $P(x)\in X$ such that $P(P(x))=x$.   Although this $P$ is first defined to act on individual elements $x\in X$ we can freely extend it to act also on sets of objects in a natural way: $P(\{x_1,x_2,\ldots\})=\{P(x_1), P(x_2),\ldots\}$. While we will mostly show $P$ acting like a function, $P(x)$ we will also sometimes write $Px$ for exactly the same expression, if space or other typesetting considerations make one more readable than the other. 
\label{def:parop}
\begin{remark}
Strictly speaking the above definition only defines an involution and does not exclude $P$ being the identity operator. The definition does not even require $P$ be anything connected to spatial inversion, even though such a connection would be very important for the measurement to be interesting from some physics perspectives!  The reasons we steer clear of specifying the physical connection between $P$ and (say) spatial inversions are: (i) the precise nature of $P$ can vary considerably from one problem to another\footnote{For geometric objects in three dimensions parity inversion may involve flipping the signs all three spatial co-ordinates $P(x,y,z)=(-x,-y,-z)$, whereas for objects in two dimensions the parity inversion may be only deemed to invert one co-ordinate $P(x,y)=(x,-y)$ since to invert both might just induce a 180-degree rotation. If there were a mirror plane at $x=3$ a more appropriate $P$ for three dimensions could be $P(x,y,z)=(6-x,y,z)$. More interestingly  it could be that $X$ is an `affine space' (i.e.~one with no real notion of `origin') so that coordinate negation \textit{per se} is not even meaningful despite the clear existence of the concept of spatial inversion image.  An example would be a situation in which $X$ were the space of `observed collisions up to rotations and translations'.}, and (ii) regardless of the physical nature of $P$, the only feature of it that is exploited by the mathematics of chiral measures is that $P$ is an involution.  It must be understood, of course, that the appropriateness of the chiral measures obtained are intimately connected to the definition of $P$ used.  If a foolish or physically irrelevant choice of $P$ (such as the identity) is used, then it should be no surprise if chiral measures turn out to be zeros.
\end{remark}
\end{definition}
\begin{definition}[Parity-even measurements]
A measurement $m$ is termed \textit{parity-even} if and only if $m(x)=m(P(x))$ for all $x\in X$.   The name \textit{parity-invariant} would be just as appropriate.
\label{def:even}
\end{definition}
\begin{definition}[Parity-odd measurements]
A measurement $m$ is termed \textit{parity-odd} if and only if $m(x)=-m(P(x))$ for all $x\in X$.\label{def:odd}
\end{definition}
\begin{lemma}
Any measurement  $m(x)$ can be written as a sum of a parity-even measurement and a parity-odd measurement.\label{lem:evenodddecomp}
\begin{proof}
It is evident from Definitions~\ref{def:parop} and \ref{def:even} that $m_e(x)\equiv \frac 1 2 (m(x)+m(P(x)))$ is parity-even.  It is evident from Definitions~\ref{def:parop} and \ref{def:odd} that $m_o(x)\equiv\frac 1 2 (m(x)-m(P(x)))$ is parity-odd.  Plainly $m(x) = m_e(x)+m_o(x)$.
\end{proof}
\end{lemma}

\subsection{Similarity under a symmetry}

\begin{remark}
Sometimes objects can have one more properties which are not affected by changes to others.   And sometimes the former properties may be the only ones people are interested in.   E.g.~a supermarket customer may only care about the colour of bananas he or she is thinking of purchasing, and the same customer may be aware that the colours of the bananas are invariant with respect to relocations or reorientations of the bananas within the shop.   Consequently it can be useful to talk about sets of objects which are identical to each other up to rotations or translation or other `unimportant' symmetries or transformations. These sets will  contain objects on which measurements (such as colour) will always agree.   Those desires motivate the next three definitions.
\end{remark}
\begin{definition}[Similar objects]
Two objects $x_1\in X$ and $x_2\in X$ are termed \textit{similar} under the symmetry group $S$ if there exists a transformation $s$ in a symmetry group $S$ acting on $X$ for which $x_1=s x_2$. Since $S$ is a group this is an equivalence relation and so is denoted $x_1 \sim x_2$.  
\end{definition}
\begin{definition}[$S(x)$ and $X/\mathord{\sim}$]
Since $\sim$ is an equivalence relation it partitions $X$ into equivalence classes.  The set of all objects similar to $x$ under $S$ is denoted $S(x)$
, while the set of \textit{all} equivalence classes into which this similarity relation partitions $X$ is denoted $X/\mathord{\sim}$.  Clearly $S(x)\in X/\mathord{\sim}$ and $S(x)\subseteq X$.  
\end{definition}
\begin{definition}[$S$-invarance]
A measurement $m$ which takes the same value on all objects which are similar to each other with respect to a symmetry group $S$ is called an \textit{$S$-invariant measurement}. \label{def:sinvariance} In other words, if a measurement $m$ is \textit{$S$-invariant}, then it will be the case that $(x\sim y ) \implies (m(x)=m(y))$, or equivalently it will be the case that $m(x)=m( s x)$ for all $x\in X$ and for all $s\in S$.
\begin{remark}
While $S$-invariant measurements are still formally measurements on objects $x\in X$, their invariance with respect to the action of elements of $S$ means that such measurements may also be thought of as functions on the equivalence classes into which $S$ partitions $X$.   Consequently we will sometimes see (slight) abuses of notation in which measurements are applied to members of $X/\mathord{\sim}$ as well as to objects in $X$.
\end{remark}
\end{definition}
\begin{definition}[Ordering a set $X$ with a relation `$<$']
A relation `$<$' is said to \label{def:ordering} ``\textit{order a set $X$}''  if it is the case that for any two elements $x,y\in X$ one and only one of the following three expressions is true: (i) $x<y$, (ii) $y<x$, (iii) $x=y$.  [Contrast with Def.~\ref{def:sinvarordering}.]
\end{definition}\noindent
\begin{remark}The definition of ordering in Def.~\ref{def:ordering} is weaker than that traditionally used to define an ordering of a field.  As such, the complex numbers (though not usually considered to be an ordered field) would qualify as being orderable by Def.~\ref{def:ordering}. For example: it would be possible to define a qualifying relation `$<$' between complex numbers $x$ and $y$ (making use of the ordinary `less than' comparison operator for reals) in the following manner:
``$x<y$ is defined to be: true if $\Re(x)<\Re(y)$,
false if $\Re(y)<\Re(x)$,
true if $\Re(x)=\Re(y)$ and $\Im(x)<\Im(y)$, and
false otherwise.''
\end{remark}
\begin{definition}[Ordering a set $X$, up to a symmetry group $S$, with a relation `$<$']
A relation `$<$' is said to  \label{def:sinvarordering} ``\textit{order a set $X$, up to a symmetry group $S$}'' if ``\textit{the set $X/\mathord{\sim}$ is ordered by `$<$'\ }''. [Contrast with Def.~\ref{def:ordering}.]
\end{definition}
\subsection{Special symmetry groups}
\begin{definition}[Special symmetry groups]
We define a \textit{special} \label{def:special} symmetry group $S$ to be a symmetry group $S$ for which
\begin{align}
P s P \in S\qquad\forall s\in S.
\end{align}
\end{definition}
\begin{lemma}
If $S$ is a \textit{special} \label{lem:PSisSPforSPECIAL} symmetry group and $x\in X$, then $P(S(x))=S(P(x))$.
\begin{proof}
Firstly: $P(S(x))\subseteq S(P(X))$ since:
\begin{align*}
\left[\vphantom{\int} a\in P(S(x)) \right] 
&\iff
\left[\vphantom{\int}
\exists s\in S\text{\ s.t.\ }a=Psx
\right]
\\
&\iff
\left[\vphantom{\int}
\exists s\in S\text{\ s.t.\ }a=PsPPx
\right]\qquad\text{(since $P^2=1$)}
\\
&\implies
\left[\vphantom{\int}
\exists s\in S\text{\ s.t.\ }a=sPx
\right]\qquad\text{(since $S$ is special)}
\\
&\iff
\left[\vphantom{\int}
a\in S(P(x))
\right].
\end{align*}
Secondly: $S(P(x))\subseteq P(S(X))$ since:
\begin{align*}
\left[\vphantom{\int} a\in S(P(x)) \right] 
&\iff
\left[\vphantom{\int}
\exists s\in S\text{\ s.t.\ }a=sPx
\right]
\\
&\iff
\left[\vphantom{\int}
\exists s\in S\text{\ s.t.\ }a=PPsPx
\right]\qquad\text{(since $P^2=1$)}
\\
&\implies
\left[\vphantom{\int}
\exists s\in S\text{\ s.t.\ }a=Psx
\right]\qquad\text{(since $S$ is special)}
\\
&\iff
\left[\vphantom{\int}
a\in P(S(x))
\right].
\end{align*}
\end{proof}
\end{lemma}
\subsection{Chirality}
\begin{definition}[Achirality]
An object $x\in X$ is defined to be \textit{achiral} with respect to the parity operator $P$ and symmetry group $S$ if and only if $S(x)=S(P(x))$. Equivalent definitions of achirality  include `$x$ is achiral if and only if $x\in S(P(x))$' and `$x$ is achiral if and only if $P(x)\in S(x)$'.  Informally this definition says that an object is achiral if it is not distinguishable from its mirror image up to an `uninteresting' transformation.  This definition of achirality is independent of any measurement process.  Later on an  (inequivalent) measurement-dependent form of achirality is introduced (see Def.~\ref{def:nonabsoluteachirality}). When it is necessary to distinguish the form defined here from the form defined there, the form defined here is sometime termed `achirality (in an absolute sense)'. This emphasises its measurement independence.
\label{def:achiral}
\end{definition}
\begin{definition}[Chirality]
An object $x$ which is not achiral is termed \textit{chiral} (with respect to the parity operator $P$ and symmetry group $S$).\label{def:chiral}  As was the case with the definition of achirality, this definition of chirality may be referred to as `chirality (in an absolute sense)' whenever it is necessary to emphasise that it is independent of any measurement process.
\end{definition}
\begin{lemma}
Any parity-odd $S$-invariant \label{lem:zerointerestingforpoddsinvariantmeasures} measurement $m$ yields zero  when applied to an achiral object $x$.  In other words $m(x)=0$, where 0 is the origin of the relevant measurement space $V$, whenever $x$ is achiral.
\begin{proof}
If $x\in X$ is achiral Def.~\ref{def:achiral} says that $P(x) \in S(x)$ and so there exists an $s\in S$ such that $P(x) = sx$.  Applying the measure $m$ to both sides yields $m(P(x)) = m(s x)$.  But $S$-invariance (Def.~\ref{def:sinvariance}) ensures that $m(s x) = m(x)$ and so $m(P(x))=m(x)$.  The desired result thus follows from $m$ being parity-odd (Def.~\ref{def:odd}).
\end{proof}
\end{lemma}
\subsection{The process of measuring is always partly convention-bound: constrained and guided by utility and goals.}
\begin{remark}
The associations between the numbers we ascribe to measurements and the physical properties being measured are sometimes so deeply ingrained in culture or convention that we may have forgotten where they came from, or that things could be otherwise.  The co-existence of Kelvin, Fahrenheit and Celsius reminds us that the location of a zero on a temperature scale is largely a matter of convention, with different choices more or less convenient to different groups of people.   But since all three scales associate hotter things with `increasingly positive' numbers, it is harder to remember that history could have given us the opposite.  Going further than this: the  presence of factors like $e^{-\frac E {kT}}$ in the Arrhenius equation or in Boltzmann factors reminds us that some forms of intelligent life may even have chosen to associate their notion of `temperature' with what we would think of as $\beta=-1/{kT}$.   (Members of the public in such civilisations might have less trouble appreciating why absolute zero is unattainable as for them it would sit at `$\beta=-\infty$'.)   The purpose of the preceding digression is to emphasise that there is considerable scope to decide what properties should be possessed by measurements which purport to measure chirality -- or indeed which purport to measure any new quantity.  What ultimately determines whether a given measurement is good or bad is whether or not it helps you achieve practical goals in real-world problems.  Clues toward which routes to take when defining new measurement systems cannot come exclusively from theorems, lemmas and proofs -- they must come from considerations relating to end-goals.

We therefore digress briefly to consider a few common questions which chiral measures could be employed in answering, so that such considerations are in mind in the section which follows.
\end{remark}

\subsection*{Interlude: five common questions relating to chirality}
The following five questions which are connected to chirality in some way, and come up in a variety of guises across many fields:
\begin{enumerate}
\item[Q1]
Is an object $x$ different to its mirror image -- i.e.~is it chiral? [yes / no]
\item[Q2]
How different is $x$ from its mirror image?
\item[Q3]
Can every chiral object $x$ be labeled  `right handed' or `left  handed' to distinguish it from its mirror image?  Or are things  more complicated than that?
\item[Q4]
If handedness can (sometimes) be more complicated than just `left vs. right', what are the best ways of describing the handedness of an object $x$ in general?
\item[Q5]
Does a given dataset provide any evidence of Parity Violation, either in the laws of physics or elsewhere?
\end{enumerate}
Of the above questions, the first two are the easiest to answer.  Although implementation issues can present challenging practical problems, in principle Q1 and Q2 could be answered by performing an optimisation (or fit) which seeks to find the minimum distance 
\begin{align}
d_\text{min}(x) &\equiv \min_{s,t,u\in S} d(s x,t P(u x))\label{eq:exampleofanonparity}
\end{align}
between a given object $x$ (or any object deemed equivalent to it under $S$) and its perturbed inverted image $P(x)$ (or any objects deemed equivalent to the parity inversions of any objects equivalent to $x$ under $S$) where $S$ is the group of `unimportant' symmetries.
Provided that the distance measure $d(x,y)$ betweeen objects $x$ and $y$ is a `semimetric' (i.e.~one which is symmetric ($d(x,y)=d(y,x)\ge 0$) and yields zero only when its inputs are identical (i.e.~$(d(x,y)=0 \implies (x=y)$) then the answer to Q1 would be `no' iff  $d_{\text{min}}(x)=0$.  Furthermore, the value $d_{\textrm{min}}$ could itself constitute an\footnote{The existence of many different possible semimetrics $d$  result in a large number of potential ways of answering Q2, for there are clearly many ways in which `closeness' between two objects can be quantified.  And, of course, there are many other methods of answering Q1 and Q2 that do not use distance metrics like that in \eqref{eq:exampleofanonparity} at all -- the distance-based approach described is just one of many possibilities. } $S$-invariant answer to Q2.

Q3 an Q4 are much harder to answer than Q1 and Q2, yet they are the real focus of this paper. The reason we pay considerable attention to them, despite the challenges they present, is that they are the only route to answering the big and interesting questions about Parity Violation (Q5). To be able to claim one has evidence for parity violation it is \textit{not sufficient} to have observed that there exist objects $x$ which are different from their mirror images, and so answers to Q1 and Q2 are not helpful.  Any bird standing on \textbf{one} leg is different from its mirror image (i.e.~is chiral) yet it is impossible to work out whether flamingos have a preference for one leg over the other by simply counting the number of legs each bird is standing on at any given time!  Evidence for parity violation (Q5) can only come from measurements which are able to distinguish objects from their mirror images (i.e.~which can measure handedness in some way).  One must determine whether more flamingos stand on their right leg than their left. These are in the domain of Q3 and Q4. 

When introducing the concept of `\textit{chiral measures}' we will therefore find it important to distinguish between `chiral measures which do not distinguish objects from their mirror images' and `chiral measures which do'. Loosely speaking  the former ones become the `\textit{parity-even chiral measures}' and the latter ones the  `\textit{parity-odd chiral measures}' of Defs.~\ref{def:parities} and \ref{def:nonparities}.

\textit{Parity-odd chiral measures} are the main focus of the paper as only they are capable of addressing questions concerning evidence for parity violation (Q5) which are likely to be most important to particle physicists.  Moreover, the proof of Theorem~\ref{thm:onlyneedoddchiralfuncs} leads to a Corollary~\ref{cor:ffhkdfffyeijwhdf} which shows that, in the circumstances a particle physicist might reasonably expect to encounter, there is no information contained within parity-even chiral measures that cannot be extracted instead from parity-odd chiral measures.   Corollary~\ref{cor:ffhkdfffyeijwhdf} (and the remark following it) therefore explains why it is the case that studies relating to Parity in particle physics can (and frequently do) exclusively focus on parity-odd chiral measures, and why (consequently) those measures deserve to be given a special name: `parities' in Section~\ref{sec:nowweareonapritys}.

\subsection{Chiral Measures\label{sec:chiralmeasures}}


\begin{definition}[Chiral-measures]
$m$ is defined to be a \textit{chiral measure} if and only if the following two requirements are satisfied:
\label{def:chiralmeasure}
\begin{enumerate}
\item
$m$ is a \textit{measurement} (see Def.~\ref{def:measurement}) which is 
$S$-invariant (see Def.~\ref{def:sinvariance}); and\label{rqt:chiralmeasureone}
\item
$m(x)=0$ whenever $x$ is  \textit{achiral} (see Def.~\ref{def:achiral}) with respect to the parity operator $P$ and symmetry group $S$.
\end{enumerate}
\end{definition}
\begin{remark}
The $S$-invariance requirement in Def.~\ref{def:chiralmeasure} is present because if it were absent then two objects which are similar under $S$ (i.e.~which only differ in ways which are supposed to be irrelevant for the purposes of assessing chirality or achirality according to Defs.~\ref{def:achiral} and \ref{def:chiral}) could have different chiralities. 
\end{remark}
\begin{remark}
The second requirement of Def.~\ref{def:chiralmeasure} is that all achiral objects are mapped to the origin of the space $V$.  This is really two sub-requirements making a single condition: (i) that every achiral object should map to a \textbf{single} location in $V$ (rather than to multiple places, or onto a sub-manifold of $V$) and (ii) that this single location shall be the \textbf{origin} of $V$.  Each of these sub-requirements deserves more discussion. Taking each in turn:

The sub-requirement that a measurement $m$ should map all achiral objects  to a \textbf{single} location  is made because an  achiral object is simply \textit{the same as its mirror image}.  A truly achiral object cannot have `more' or `less' chirality than any other truly achiral object.  A truly achiral object can of course be bigger or smaller or hotter or colder than another -- but such differences are  are manifestations of its non-chiral properties.  The desire to map achirality to a single location is therefore tied to the fact that `being achiral' (as opposed to being chiral) is `something without nuance'.  There is simply only one way of being achiral, and so any measurement (if it is to be a measurement of chirality rather than of other properties) should reflect this.

The sub-requirement that the single destination of all achiral objects be the \textbf{origin} of $V$ (rather than somewhere else in $V$) is partly convention and partly forced upon us.  The forced element comes from Lemma~\ref{lem:zerointerestingforpoddsinvariantmeasures} which says that any parity-odd $S$-invariant measurement unavoidably maps all achiral objects to the origin.  Since chiral measures are (by definition) $S$-invariant,  we see that there is no-choice about where to put the  fixed point for any parity-odd chiral measures.  In principle parity-even chiral measures could have their fixed achiral points placed somewhere other than the origin -- however any measurement making such a choice would be \textit{equivalent} (in the sense of Def.~\ref{eq:equivalentmeasurements}) to another using the origin as its fixed achiral point. Consequently, no generality is lost by asking all achiral measurements to map to 0.  Finally, many simplifications and benefits are afforded by this convention.  For example: it is much easier to build chiral measures from other ones (with inner products, outer products and linear combinations, \textit{etc}.) if the the fixed achiral point is the origin of $V$.
\end{remark}

\subsection{Parity-Even and Parity-Odd Chiral Measures\label{sec:parityvsnonparity}}

Lemma~\ref{lem:evenodddecomp} tells us that every \textit{measurement} is the sum of a \textit{parity-even} measurement (Def.~\ref{def:even}) and a \textit{parity-odd} one (Def.~\ref{def:odd}).  Since a \textit{chiral measure} is just a particular kind of measurement (see definition~\ref{def:chiralmeasure}) it follows that every chiral measure has a  \textit{parity-even} and a \textit{parity-odd} part.  We give special names to those parts:
\begin{definition}[Parity-Odd Chiral Measures]
The parity-odd parts of \label{def:parities} chiral measures are termed `\textit{parity-odd chiral measures}'.  The name is not an abuse of notation because such things may be verified to be \textit{chiral measures} in their own right.
\end{definition}
\begin{definition}[Parity-Even Chiral Measures]
The parity-even parts of\label{def:nonparities} chiral measures are termed `\textit{parity-even chiral Mmeasures}'.  The name is not an abuse of notation because such things may be verified to be \textit{chiral measures} in their own right. [An example of an even chiral measure could be the distance $d_\text{min}(x)$ of Eq.~\eqref{eq:exampleofanonparity}.]
\end{definition}
\begin{corollary}
Every \textit{chiral measure} is a sum of a \textit{parity-even chiral measure} and a \textit{parity-odd chiral measure}. [This is a simple consequence of Lemma~\ref{lem:evenodddecomp} taken together with Definitions~\ref{def:parities} and \ref{def:nonparities}.]
\end{corollary}

\begin{lemma}
Every one-dimensional complex \textit{parity-even chiral measure} $m_{1\mathbb{C}}$ is \textit{equivalent} to a two-dimensional real \textit{parity-even chiral measure} $m_{2\mathbb{R}}$.  Moreover, if $m_{1\mathbb{C}}$ is continuous on $X$ then so is $m_{2\mathbb{R}}$.\label{lem:ghfsewsddcv} 
\begin{proof}
If $m_{1\mathbb{C}} : x\in X \mapsto m_{1\mathbb{C}}(x) \in \mathbb{C}$ then \textbf{define} $m_{2\mathbb{R}}$ to be such that   $m_{2\mathbb{R}}: x\mapsto (\Re(m_{1\mathbb{C}}(x)),\Im(m_{1\mathbb{C}}(x)))\in\mathbb{R}^2$.  This $m_{2\mathbb{R}}$ has the desired properties.
\end{proof}
\end{lemma}
\begin{lemma}
Every one-dimensional real \textit{parity-even chiral measure} $m_{1\mathbb{R}}$ is \textit{equivalent} to a two-dimensional real \textit{parity-even chiral measure} $m_{2\mathbb{R}+}$ for which $m_{2\mathbb{R}+}=|m_{2\mathbb{R}+}|$.  Moreover, if $m_{1\mathbb{R}}$ is continuous on $X$ then so is $m_{2\mathbb{R}+}$.\label{lem:gmmnmnbnvgfghgf}
\begin{proof}
If $m_{1\mathbb{R}} : x\in X \mapsto m_{1\mathbb{R}}(x) \in \mathbb{R}$ then \textbf{define} $m_{2\mathbb{R}+}$ to be such that  $m_{2\mathbb{R}+}:  X \rightarrow \mathbb{R}^2$ via $m_{2\mathbb{R}+} : x\mapsto (\max[m_{1\mathbb{R}}(x),0],\max[-m_{1\mathbb{R}}(x),0])$.  This $m_{2\mathbb{R}+}$ has the desired properties.
\end{proof}
\end{lemma}
\begin{corollary}
Every one-dimensional complex \textit{parity-even chiral measure} $m_{1\mathbb{C}}$ is \textit{equivalent} to a four-dimensional real \textit{parity-even chiral measure} $m_{4\mathbb{R}+}$ for which $m_{4\mathbb{R}+}=|m_{4\mathbb{R}+}|$.  \label{cor:trfreedfgbmnbjd} Moreover, if $m_{1\mathbb{C}}$ is continuous on $X$ then so is $m_{4\mathbb{R}+}$.
\begin{proof}
The complex \textit{parity-even chiral measure} $m_{1\mathbb{C}}$ may be converted to an intermediate two-dimensional real \textit{parity-even chiral measure} $m_{2\mathbb R}$ using Lemma~\ref{lem:ghfsewsddcv}. Subsequently, each of those real dimensions may be doubled into non-negative real dimensions by the use of Lemma~\ref{lem:gmmnmnbnvgfghgf}, making four dimensions in total.
\end{proof}
\end{corollary}
\begin{corollary}
Every \textit{parity-even chiral measure} $m$ is \textit{equivalent} to a purely real \textit{parity-even chiral measure} $m_{n\mathbb{R}+}$ for which  \label{cor:poihwkfdvdofrhi} $m_{n\mathbb{R}+}=|m_{n\mathbb{R}+}|$. Moreover, if $m$ is continuous on $X$ then so is $m_{n\mathbb{R}+}$.
\begin{proof}
The most general measure has a co-domain of the form $V=\mathbb{R}^a \otimes \mathbb{C}^b$ for some non-negative integers $a$ and $b$.  By the use of $a$ instances of Lemma~\ref{lem:gmmnmnbnvgfghgf} and $b$ instances of Corollary~\ref{cor:trfreedfgbmnbjd} is may be observed that a continuous mapping into the positive parts of $\mathbb{R}^{2a+4b}$ is possible.
\end{proof}
\end{corollary}

\begin{theorem}
\label{thm:onlyneedoddchiralfuncs}
Provided that there exists a relation `$<$' which is able to \textit{order $X$, up to a \textbf{special} symmetry group $S$},\footnote{See Defs.~\ref{def:sinvarordering} and \ref{def:special}.} then every \textit{parity-\textbf{even} chiral measure} $m_e$ is \textit{equivalent} to  the \textit{absolute form}\footnote{See Def.~\ref{def:absoluteformof}.} $|m_o|$ of a real \textit{parity-\textbf{odd} chiral measure} $m_o$.
\begin{proof}
Using Corollary~\ref{cor:poihwkfdvdofrhi} one may construct from $m_e$ an \textit{equivalent} real \textit{parity-\textbf{even} chiral measure} $m_{n\mathbb{R}+}$ for which $m_{n\mathbb{R}+}=|m_{n\mathbb{R}+}|$.  One may then construct from $m_{n\mathbb{R}+}$ a real \textit{
measure} $m_o$ according to:
\begin{align}
m_o(x) &= 
\begin{cases}
+m_{n\mathbb{R}+}(x) &\text{if $S(x)<S(P(x))$,} \\
-m_{n\mathbb{R}+}(x) &\text{otherwise.}
\end{cases}\label{eq:modef}
\end{align}
Furthermore, $m_o$ is evidently a real \textit{chiral measure} 
because: (i) $m_o$ is $S$-invariant:
\begin{align*}
m_o(sx) &=
\begin{cases}
+m_{n\mathbb{R}+}(sx) &\text{if $S(sx)<S(P(sx))$,} \\
-m_{n\mathbb{R}+}(sx) &\text{otherwise}
\end{cases}
\\
&= 
\begin{cases}
+m_{n\mathbb{R}+}(x) &\text{if $S(x)<S(P(sx))$,} \\
-m_{n\mathbb{R}+}(x) &\text{otherwise}
\end{cases}
\\
&= 
\begin{cases}
+m_{n\mathbb{R}+}(x) &\text{if $S(x)<P(S(sx))$,} \\
-m_{n\mathbb{R}+}(x) &\text{otherwise}
\end{cases}\qquad\text{(Lem.~\ref{lem:PSisSPforSPECIAL} and $S$ `special')}
\\
&= 
\begin{cases}
+m_{n\mathbb{R}+}(x) &\text{if $S(x)<P(S(x))$,} \\
-m_{n\mathbb{R}+}(x) &\text{otherwise}
\end{cases}
\\
&= 
\begin{cases}
+m_{n\mathbb{R}+}(x) &\text{if $S(x)<S(P(x))$,} \\
-m_{n\mathbb{R}+}(x) &\text{otherwise}
\end{cases}\qquad\text{(Lem.~\ref{lem:PSisSPforSPECIAL} and $S$ `special')}
\\
&= m_o(x)
\end{align*}
and (ii) $m_o(x)$ is zero for any achiral $x\in X$ (this is because $m_{n\mathbb{R}+}(x)$ is itself zero for any such $x$ on account of being itself a chiral measure).
Furthermore $m_o(x)$ is parity-odd (making it an real \textit{parity-\textbf{odd} chiral measure}) because:
\begin{align*}
m_o(P(x)) &=
\begin{cases}
+m_{n\mathbb{R}+}(P(x)) &\text{if $S(P(x))<S(P(P(x)))$,} \\
-m_{n\mathbb{R}+}(P(x)) &\text{otherwise}
\end{cases}
\\
&= 
\begin{cases}
+m_{n\mathbb{R}+}(x) &\text{if $S(P(x))<S(x)$,} \\
-m_{n\mathbb{R}+}(x) &\text{otherwise}
\end{cases}\quad\text{($m_{n\mathbb{R}+}$ parity-even and $P^2=1$)}
\\
&= 
\begin{cases}
+m_{n\mathbb{R}+}(x) &\text{if $S(P(x))<S(x)$,} \\
-m_{n\mathbb{R}+}(x) &\text{if $S(x)<S(P(x))$,}\\
-m_{n\mathbb{R}+}(x) &\text{if $S(x)=S(P(x))$}
\end{cases}\quad\text{(trichotomy in Def.~\ref{def:ordering})}
\\
&= 
\begin{cases}
+m_{n\mathbb{R}+}(x) &\text{if $S(P(x))<S(x)$,} \\
-m_{n\mathbb{R}+}(x) &\text{if $S(x)<S(P(x))$,}\\
0 &\text{if $S(x)=S(P(x))$}
\end{cases}\quad\text{($m_{n\mathbb{R}+}$ a chiral measure)}
\\
&= 
\begin{cases}
+m_{n\mathbb{R}+}(x) &\text{if $S(P(x))<S(x)$,} \\
-m_{n\mathbb{R}+}(x) &\text{if $S(x)<S(P(x))$,}\\
+m_{n\mathbb{R}+}(x) &\text{if $S(x)=S(P(x))$}
\end{cases}\quad\text{($m_{n\mathbb{R}+}$ a chiral measure)}
\\
&= 
\begin{cases}
-m_{n\mathbb{R}+}(x) &\text{if $S(x)<S(P(x))$,}\\
+m_{n\mathbb{R}+}(x) &\text{otherwise}
\end{cases}\\
&=
-m_o(x).
\end{align*}
The proof concludes by noting that $|m_o|(x) = m_{n\mathbb{R}+}(x)$ since  $m_{n\mathbb{R}+}=|m_{n\mathbb{R}+}|$.
\end{proof}
\end{theorem}
\begin{corollary}[Parity-even chiral measures serve no useful purpose]
Provided that there exists a relation `$<$' which is able to \textit{order $X$, up to a \textbf{special} symmetry group $S$}, a consequence of Theorem~\ref{thm:onlyneedoddchiralfuncs} is that an arbitrary \textit{chiral measure} $m$ (whether parity-even, or parity-odd, or a mixture) \label{cor:ffhkdfffyeijwhdf} contains no information which cannot be extracted from an exclusively \textit{\textbf{parity-odd} chiral measure}.
\end{corollary}
\begin{remark}
Corollary~\ref{cor:ffhkdfffyeijwhdf} was (somewhat provocatively)  subtitled `parity-even chiral measures serve \textit{no useful purpose}'. These words deserve some clarification.  Clearly parity-even chiral measures \textit{do} serve useful purposes as they are used in many fields.\footnote{The ten pages of Section 2 of \cite{PetitjeanReview} (``A Transdisciplinary review of Chirality and Symmetry Measures'') provide succinct descriptions of more than one hundred ``Chriality Measures in Chemistry and Physics''.  A very large fraction of these would be considered \textit{parity-even chrial measures} by Def.~\ref{def:nonparities}.}
The words `no useful purpose' are not intended to undermine those practices.  Instead our focus in making the statement is to emphasise two things: (i) that the criterion requiring $X$ to be orderable up to a special symmetry group $S$ is almost always trivially satisfied in any context likely to be encountered by a reader of this paper,\!\footnote{Any objects which can be represented in a computer can be reduced to ones and zeros. Ones and zeros can be converted into a bigger numbers which can then be ranked to create a numerical ordering on the objects.  Any context in which one can imagine `storing an element of $X/\mathord{\sim}$ in a  computer' is therefore one for which Cor.~\ref{cor:ffhkdfffyeijwhdf} applies. It is hard to imagine situations in which objects in $X/\mathord{\sim}$ could not be stored (in principle) on a computer.  If (say) the symmetry group $S$  contained all proper Lorentz boosts, rotations and translations of collision events $x$ in a collider, then trivially the equivalence classes in $X/\mathord{\sim}$ could be indexed (i.e.~stored in a computer) by transforming such elements into a canonical form using elements of $S$.  E.g.~a canonical transformation might be `boost $x$ to its rest frame, then translate $x$ until its centre of mass is at the origin, then rotate $x$ until the most energetic particle is along the $z$-axis, \ldots'.  The kinds of $x$ resulting from the above procedure would then uniquely index the equivalence classes in $X/\mathord{\sim}$, and they are also evidently storable, and so are also orderable.}  and (ii) that  therefore all work on questions concerning parity can (without any loss of generality) be conducted using only \textit{\textbf{odd} chiral measures}, at least in principle.  Whether a given analysis will actually see a practical benefit from discarding parity-even chiral measures will inevitably depend on the specifics of that particular analysis, and that is why so many different chiral measures co-exist. It could be that being able to focus on a single type of chiral measure (rather than two) provides simplifications. Or, if the parity-odd measures for a given problem are more complex or more unwieldy than the even measures for the same problem, then the reverse might be true. Nonetheless, the important point being made is that \textit{in principle} no information is lost by an exclusive focus on parity-odd chiral measures. Knowing this allows us to make progress.
\end{remark}
\begin{lemma}
The parity-odd chiral measure $m_o$ in Eq.~\eqref{eq:modef} need not be continuous,  even if the parity-even chiral measure $m_e$ (in Theorem~\ref{thm:onlyneedoddchiralfuncs}) from which it was generated is itself continuous.  
\label{lem:monotnecessarilycontinuous}
\begin{proof}
Consider, the star domain object shown in Figures~\ref{fig:blobofdeath} and \ref{fig:slideofdeath}.  Those figures describe a continuous path $x(t)$ through a space of objects (star-domains, defined later)  controlled by a parameter $t\in[0,1]$, and (by construction) $x(0)$ and $x(1)$ are mirror images of each other.  If $m_e$ were to be defined to be the sum of the squares of the  quantities $I_{1,2;1}$ and $I_{3,4;2}$ plotted in Figure~\ref{fig:slideofdeath} and defined in Eq.~\eqref{eq:formulaforcorrelatorinvariant}, then no choice of `$<$' could unpack $m_e$ into a continuous $m_o$, even though  $m(x(t))=(I_{1,2;1}(x(t)), I_{3,4;2}(x(t)))\in \mathbb{R}^2$ is trivially a parity-odd chiral measure from which $m_e(x(t))$ could be computed for any $t$. The reason the construction used in Theorem~\ref{thm:onlyneedoddchiralfuncs} cannot generate a continuous $m_o$ for this particular $m_e$ is as follows: 
\begin{enumerate}
\item
this $m_e(x(t))$ is real and one-dimensional and so the construction method described will first generate a $m_{n\mathbb{R}+}$ that is two dimensional and has $m_{n\mathbb{R}+}=\left|m_{n\mathbb{R}+}\right|$; 
\item
because this $m(x(t))$ is never negative, the second component of $m_{n\mathbb{R}+}$ will always be zero; more specifically we will have $m_{n\mathbb{R}+}(x(t))=(m_e(x(t)),0)$; 
\item
consequently the construction will require that $m_o(x(t))=(u_{\mathord{<}}(t) m_e(x(t)), 0)$ where $u_{\mathord{<}}(t)$ is a sign ($\pm 1$) which depends on the value of $t$ and the relation `$<$'; 
\item
as $x(0)$ is (by design) the parity inversion of  $x(1)$ it must be the case that $m_o(x(0))=-m_o(x(1))$ and so $m_o(x(t))$ must take the value $(+K,0)$ at $t=0$ and must reach the value $(-K,0)$ at $t=1$, for some non-zero $K\in\mathbb{R}$; 
\item
if $m_o(x(t))$ were to get from $(+K,0)$ to $(-K,0)$ continuously it would have to go via $(0,0)$ since the second component if $m_o(x(t))$ is not able to change;
\item
however $m_o(x(t))$ cannot equal $(0,0)$ for any $0<t<1$ since $|u_{\mathord{<}}(t)|=1$ and $m_e(x(t))$ is bounded below by a positive real number for all $t$ (as shown in Figure~\ref{fig:slideofdeath}).
Since this contradiction is present for any $u_{\mathord{<}}(t)$, we see that no choice of `$<$' can lead to a continuous $m_o$ by this method of construction.
\end{enumerate}
\end{proof}
\end{lemma}
\begin{remark}
This does not mean, however, that there are no continuous parity-odd chiral measures which encapsulate the information content of $m_e$. All it means is that the particular method of construction described in Theorem~\ref{thm:onlyneedoddchiralfuncs} is not always capable of delivering  \textit{continuous} chiral measures $m_o$ even though it does always deliver parity-\textit{odd} ones.
\end{remark}
\begin{remark}
Theorem~\ref{thm:onlyneedoddchiralfuncs} and Corollary~\ref{cor:ffhkdfffyeijwhdf} are in some tension with a strong statement made in Ref.~\cite{mislowweinberg}.  The conclusion of a piece of analysis therein is reported as follows\footnote{We reproduce the emphasis used in \cite{mislowweinberg}.}:
\begin{quote}
 It follows that, as a rule, \textit{continuous pseudoscalar functions cannot be used as chirality measures in three and higher dimensions.} This conclusion renders questionable current and past attempts to formulate ``chirality measures'' on the basis of pseudoscalar functions.
\end{quote}
Strictly, on account of at least two technicalities, there is no formal disagreement between \cite{mislowweinberg} and our work because:
\begin{enumerate}
\item
their statement refers to \textbf{continuous} pseudoscalars, and Lemma~\ref{lem:monotnecessarilycontinuous} has already demonstrated that the $m_o$ constructed by the algorithm of Theorem~\ref{thm:onlyneedoddchiralfuncs} need not always be continuous; and
\item
their statement refers to continuous \textbf{pseudoscalars}, which is their word for something very similar to what we have called parity-odd chiral measures -- but limited by $V$ being  a one-dimensional vector space, $\mathrm{dim}(V)=1$, rather than the more general vector space, $\mathrm{dim}(V)\ge 1$, which we permit for measurements in Def.~\ref{def:measurement}.\footnote{Do not confuse the dimension of the measurement space $V$ being mentioned in this sentence with the dimensions of objects in the input space $X$.  These are different things.  When the \cite{mislowweinberg} quotation talks of `three and higher dimensions' they are imagining molecules in three and higher dimensions, not measurements which record three or more numbers.}
\end{enumerate}
Nonetheless, despite the formal lack of disagreement, tension still exists. The two main causes of this tension are:
\begin{enumerate}
\item
Our contention that it would be a mistake to  restrict oneself to only ever considering \textbf{one}-dimensional (pseudo)scalar measurements.  While it is true (for the reasons given in \cite{mislowweinberg}) that a focus on \textbf{one}-dimensional pseudoscalars measurements would indeed show them to have limited uses, we contend that the limitations derives from the focus on the `scalar' in `pseudoscalar' not on the `pseudo'.  One would not, for example, expect to record much information about a sound by reducing it to a single number. Instead one would capture all its content in a Fourier transform or frequency spectrum consisting of many numbers. One would not attempt to quantify the location of a point in three-space with a single number -- one would use three co-ordinates $x$, $y$ and $z$.  Quantifying chirality should be no different.
\item
Secondly, even though the construction used in Theorem~\ref{thm:onlyneedoddchiralfuncs} is not always capable of generating a \textbf{continuous} $m_o$ from a continuous even measure $m_e$, we have strong reason to conjecture that the proof of Theorem~\ref{thm:onlyneedoddchiralfuncs} can be extended by the addition of a more complex construction which \textit{is} able to deliver \textbf{continuous} $m_o$ in a very wide variety of circumstances.  In the present paper this conjecture is not  well posed since the precise nature of the restrictions on $X$, $V$, $S$, $P$ and $m_e$ needed to make it concrete (and thus provable or disprovable) has not been provided.  However, a key fact is that the conjecture would undoubtedly be false if we considered only measurements for which $dim(V)=1$. 
\end{enumerate}
\subsection{Summary}
\begin{itemize}
\item
Theorem~\ref{thm:onlyneedoddchiralfuncs} has proved that (without loss of generality) one need only consider parity-odd chiral measures when performing measurements relating to parity or chirality when objects in $X$ are orderable up to a special symmetry group $S$.
\item
The above result is valuable as it provides a basis on which to base theorems about completeness for future analyses investigating issues connected with parity.
\item
The authors intend to extend Theorem~\ref{thm:onlyneedoddchiralfuncs} in a later work to a  stronger result.  It is expected that (without loss of generality) it will be possible to claim that one need only consider continuous parity-odd chiral measures when performing measurements relating to parity or chirality.  The circumstances in which this claim is expected to be true are not yet specified, but they are not expected to be significantly stronger than those already forming the preconditions for Theorem~\ref{thm:onlyneedoddchiralfuncs}.
\item
Theorem~\ref{thm:onlyneedoddchiralfuncs} (and its conjectured extension above) are not in direct conflict with statements made in \cite{mislowweinberg}. However tensions do exist in that we wish to push back against the subtext pushed by  \cite{mislowweinberg} that continuous pseudoscalars are misguided direction in which to direct efforts.  Our claim is that the parity-odd variables in sufficiently large vector spaces are, in some sense, more fundamental than their parity-even counterparts,  since the former can answer some questions that the latter cannot (such as `Is this a left hand rather than a right one?') while the reverse is not the case.
\end{itemize}
\end{remark}

\section{Parities as measures of chirality}
\label{sec:nowweareonapritys}

Given the pre-eminent role played by parity-odd chiral measures and established by the earlier parts of this document, it makes sense to give them a special name.  
Not only does a shortened name allow us to refer to parity-even chiral measures more succinctly, it will also allow us to  better connect to the common usage seen in particle physics literature.  Accordingly, we make the following definition:  
\begin{definition}[A parity, or the parity of \ldots]
On account of its importance as a fundamental concept, we refer to `\textit{a parity-odd chiral measure}' as `\textit{a parity}'. Furthermore, if $f$ is
\textit{a parity} then the quantity $f(x)$ is referred to as \textit{the parity of $x$ under $f$}.\label{def:aparity} 
\end{definition}
\begin{remark}Def.~\ref{def:aparity} comes close to systematising common terminology seen elsewhere. For example, the particle data booklet attaches parities of +1 or -1 to each of the subatomic particles in its listings.  Those numbers are (in our terms) the values $f(x)$ obtained when `a parity' $f$ (a.k.a.~`a parity-odd chiral measure') is evaluated on a species of particle $x\in\{e^+, e^-, \mu^+, \mu^-, \ldots \}$.
\end{remark}
\begin{remark}A consequence of Def.~\ref{def:aparity}  is that the word `parity' is associated with four different but related concepts:
\begin{enumerate}
\item
`the parity symmetry' is a general concept, leading to adjectival usage in expressions like `parity-even' and `parity-odd' (Defs.~\ref{def:even} and \ref{def:odd});
\item
`a parity operator' (see Def.~\ref{def:parop}) might refer to any one of the operators $P$ which is able to implement the parity symmetry up to a symmetry group $S$;
\item
`a parity' can be used (Def.~\ref{def:aparity}) as a short-hand for `a parity-odd chiral measure'; and
\item
the value $f(x)$ of a parity $f$ when evaluated on an object $x$ can be referred to as `the parity of $x$ under $f$' (also Def.~\ref{def:aparity}).
\end{enumerate}
Fortunately confusion ought not to be caused by this re-use (or over-use) of the word `parity', as it is should clear from context which usage is intended.
\end{remark}

\begin{definition}[The set of all parities]
$F_X$ is defined to be the set of all parity functions $f$ which can be defined on the set $X$.
\end{definition}
\begin{definition}[Chirality and achirality with respect to a parity $f$]
 An object $x\in X$  is said to be `\textit{achiral with respect to $f$}' if $f\in F_X$ and $f(x)=0$.  Any object in $X$ which is not \textit{achiral with respect to $f$} is said to be `\textit{chiral with respect to $f$}' and will have $f(x)\ne\textbf{0}$.  \label{def:nonabsoluteachirality} [ These definitions should not be confused with those of (absolute) achirality and chirality seen in Defs.~\ref{def:achiral} and \ref{def:chiral}. ]
\end{definition}

\begin{definition}[Absolute chirality]
An object $x\in X$ is said to be \textit{achiral} (in an absolute sense) 
if $f$ is \textit{achiral with respect to \textbf{every} $f\in F_X$.}
Any object in $X$ which is not \textit{achiral} (in an absolute sense) is said to be \textit{chiral} (in an absolute sense) and will have $f(x)\ne\textbf{0}$ for \textbf{at least one} $f\in F_X$.
\end{definition}

\begin{definition}[Scalar and Vector parities]
If the vector space $V$ into which  $f$ maps objects $x$ is also field (e.g.~$V=\mathbb{R}$ or $V=\mathbb{C}$) then we name $f$ a `\textit{scalar parity}'.  
If, instead, $f$ maps objects $x$ into a more complex (and possibly infinite dimensional) vector space (e.g.~$V=\mathbb{R}^4$ or $V=\mathbb{C}^{5}$) then we refer to $f$ as a `\textit{vector parity}'.\footnote{This naming structure does not carry with it any of the tensorial connotations that are sometimes associated with the words `scalar' and `vector'.}
\end{definition}

\subsection{Continuous parities}
 If the space $X$ is such that it is possible to perturb some object $x\in X$ by arbitrarily small amounts so as to make other arbitrarily `close-by' objects $y\in X$, then it seems not unreasonable to focus on parities $f$  which are continuous functions at that location $x$. If a parity $f$ is continuous at all such locations $x$ we call it a  `\textit{continuous parity}'. It is possible that future work will require a more structured definition of `continuity' than that just given. For the moment, however, we leave the precise definition of `continuity' to the imagination of the reader as there are more important issues to focus on first, and nothing in the text which follows requires a rigorous definition of continuity.
 
However, the main motivation\footnote{A secondary motivation is that external users of parities (e.g.~neural nets and other machine learning algorithms) may prefer to see similar objects described by similar numbers at their inputs.  Or put another way: one does not usually apply discontinuous but invertible `scrambling' functions to images  before attempting to do image recognition on them. Doing so would just make life harder!} for requesting continuity is a desire to make parity measurements in real experiments well defined: ideally  uncertainties in observational measurements of $x$, if sufficiently controlled, should lead to correspondingly small and well controlled uncertainties on the measured parity, $f(x)$.   Without continuity, reliable parity determinations would require $x$ to be measured with an impossible infinite precision.  A counter argument could  be made that a lesser requirement of \textit{piecewise} continuity would be sufficient to accomplish the goal of `reliably measurable parities' (in all but the measure-zero set of edge-cases at the boundaries). Nonetheless, motivated by  examples of places where continuous descriptions are more useful than simpler discontinuous ones\footnote{For example: software packages which make considerable use of 3D-rotations tend to represent those rotations with four-component quaternions, rather than with three Euler angles. The practical benefits of the former's continuous parameterization (notably lack of gimbal-lock and ease of composition) far outweigh the cost of an extra parameter they carry, despite the steeper learning curve.}, we will not explore \textit{piecewise} continuity in this paper.


\subsection{`Ideal' parities}

Although achiral objects $x$ always have zero parity with respect to any $f$ (this is one of the defining requirements of chiral measures) the converse is not true: an object $x$ need not be achiral (in the absolute sense) just because $x$ is achiral with respect to a particular parity $f$.  For example, the function $f$ mapping everything to zero is a (trivial) parity despite having every chiral and achiral object in its kernel.    We define an `\textit{ideal parity}' to be one that has no such exceptions:
\begin{definition}[Ideal parity]
An `\textit{ideal parity'} for a set $X$ is a parity $f$ such that  $f(x)=\textbf{0}$ if and only if $x\in X$ is achiral (in the absolute sense of Def.~\ref{def:achiral}).
\end{definition} 
\begin{corollary}
An `\textit{ideal parity'} for a set $X$ is a parity $f$ such that  $f(x)\ne\textbf{0}$ if and only if $x\in X$ is chiral (in the absolute sense of Def.~\ref{def:chiral}).
\end{corollary}

The motivation for calling such parities `ideal' is that, if you can find one, and if your objective is to be able to identify whether or not any given object $x$ is or is not chiral in an absolute sense, then in principle you don't need to find or use any other parities.  The one you have will suffice. It is `ideal' for your purposes.

\subsection{Continuous ideal scalar parities}
It would be nice if, for any set  of objects $S$, there existed at least one continuous scalar ideal parity $f$.   If it existed, such an $f$ could be declared to be `the' parity of every object in $S$. It would map any achiral object $x\in S$ to zero, and any chiral object to a non-zero real value whose sign would distinguish its left and right handed forms.  Alas, Lemma~\ref{lem:nouniversalparity} proves that there exist sets $S$ for which no such function exists.
\begin{figure}
\begin{center}
  \includegraphics[width=0.99\linewidth]{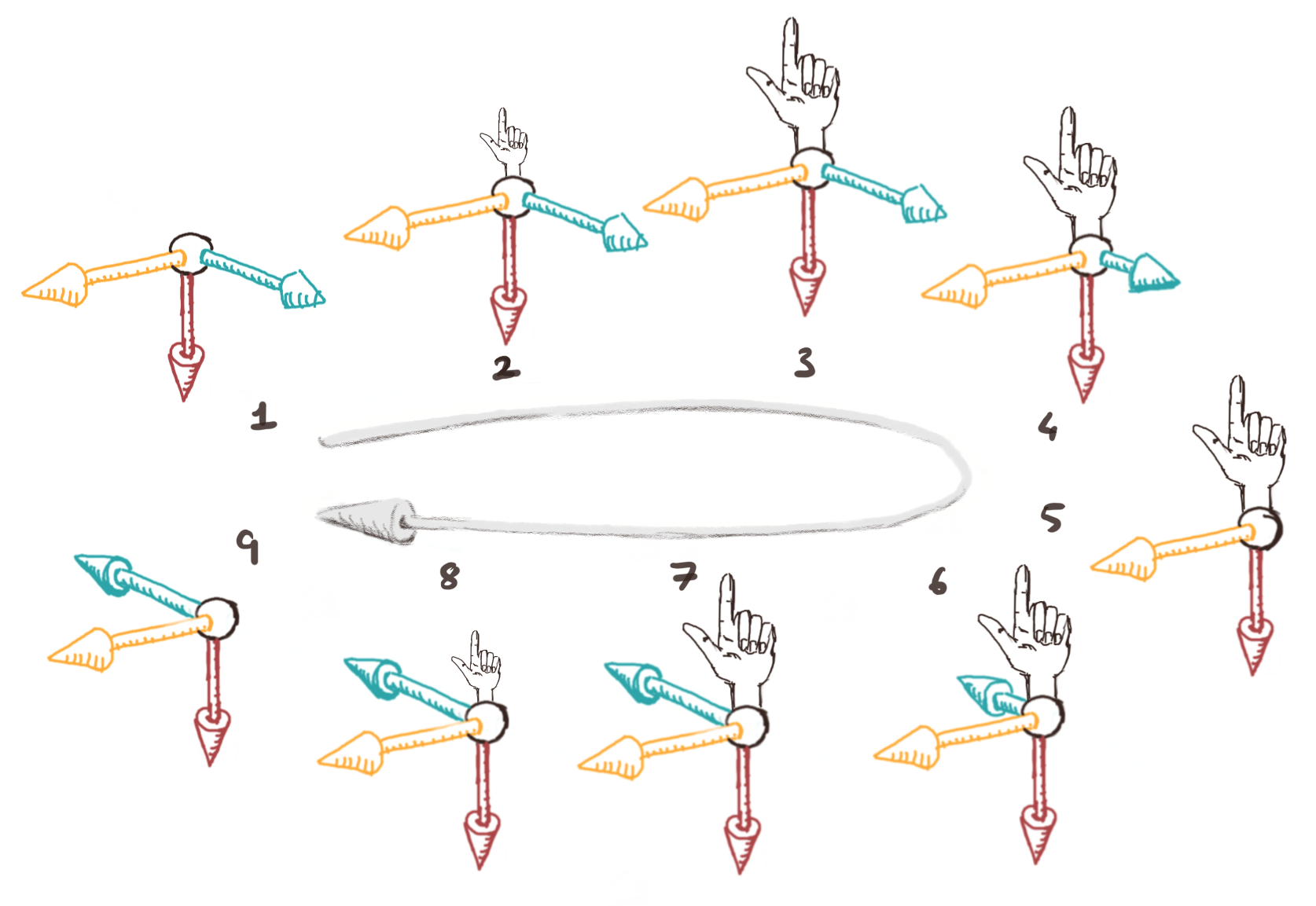}
    \caption{
    A continuous transformation of a chiral object into to a mirror image of itself.  This transformation does not pass through an intermediate achiral state and forms part of the proof of Lemma~\ref{lem:nouniversalparity}.  A more concrete formulation of this figure can be found in Figure~\ref{fig:blobofdeath}.   Conceptually the same process (in the context of molecular chemistry) is illustrated in  Figure 6 of \cite{RevModPhys.71.1745}.\label{fig:handofdeath}
  }
  \end{center}
\end{figure}
\begin{lemma}
Any sufficiently complex set $S$ of continuously deformable objects will not admit a continuous ideal scalar parity.\label{lem:nouniversalparity}
\begin{proof}
  Take $S$ to be the set of `pliable multicoloured simply-connected clay models in three dimensions'.   Suppose that for this set $S$ there exists a  continuous ideal scalar parity $f$.  
  Since some clay models are chiral, and since every clay model has a mirror image, it must be possible to find two objects in $S$ which are assigned parities of different signs by $f$. I.e.~we can assume that there exist objects $x_+$ and $x_-$ in this set such that $f(x_+)>0$ and $f(x_-)<0$.  There are many `routes' $R$ via which one could continuously deform object $x_+$ into object $x_-$ by manipulation of the clay. For example, a route `$R_1$' might first deform $x_+$ into a sphere of volume $V$ before  `un-deforming' the sphere into the shape of $x_-$. A different route $R_2$ might go via a small model of Michelangelo's David, rather than via a sphere. For any given route $R$ the shape deformation process may be parameterized using a continuous function $x_R(t)$ with parameter $t\in[0,1]$ and with boundary conditions $x_R(0)=x_+$ and $x_R(1)=x_-$.  A parameterization via a route $R$ thus induces a function $F_R(t)$ from $[0,1]$ to the reals according to the definition: $F_R(t)=f(x_R(t))$.  From our assumption about the continuity of 
$f$, the functions $F_R(t)$ would inherit continuity for themselves. Since each $F_R(t)$ is continuous and takes a different sign at each of its endpoints  ($F_R(0)=f(x_+)>0$ and $F_R(1)=f(x_-)<0$) there must exist an intermediate value of $t$ for each route (call it $t_R\in(0,1)$) at which $F_R(t_R)=0$. Furthermore,  our assumption that $f$ is an `ideal' parity tells us that the corresponding intermediate objects $x_R(t_R)$ are necessarily achiral. In other words, the existence of a  continuous ideal scalar parity would imply that: \textit{every continuous deformation of a coloured clay model from a state with positive parity to one with negative parity would necessarily have to pass through at least one achiral intermediate state}.  The sphere of the `route $R_1$' mentioned earlier is an example of such an achiral intermediate state.  However, the transformation process shown in  Figure~\ref{fig:handofdeath} (a more concrete  illustration of the same sort of transformation may be found in Figure~\ref{fig:blobofdeath}) is an example of one which \textbf{contradicts} the result just derived. It shows that it is possible to continuously deform an object of coloured clay into a mirror image of itself \textbf{without} passing through an intermediate achiral state. The trick used (creating a chiral appendage unrelated to the $x_+$ and $x_-$, then transforming the $x_+$ to $x_-$ in any way desired, before finally shrinking the chiral appendage) does not even require $x_+$ to be a mirror image of $x_-$. The existence of at least one such always-chiral path in $S$ between at least two objects of opposite parity proves that $S$ is a set for which no continuous scalar parity exists. 
\end{proof}
\end{lemma}
It should be clear that the set $S$ of coloured clay models which was chosen to prove Lemma~\ref{lem:nouniversalparity} was not particularly special.  The non-existence of a continuous ideal scalar parity is therefore a generic property of any sufficiently complex set $S$.


\subsection{Continuous ideal vector parities}
Just because typical sets $S$ of sufficient complexity lack continuous ideal \textbf{scalar} parities does not mean that those sets lack continuous ideal parities.
It just means that one must be prepared to look for continuous ideal \textbf{vector} parities.  One can imagine that in circumstances in which the objects in $S$ have only finitely many degrees of freedom, that an iterative process could be used to generate an ideal continuous vector parity for that set.  The algorithm would have steps resembling these:
\begin{enumerate}
\item
Create an (initially empty) ordered list $\vec v$ in which scalar parities will later be stored.  Also, let $T$ be a copy of the set $S$.
\item
If $T$ is empty or every element in $T$ is achiral, then go to step 5. 
\item
Find a (not-necessarily-ideal) continuous scalar parity, $p(x) : S\rightarrow \mathbb{R}$,  which gives at least one element $x\in T$ a non-zero parity.
Append this $p(x)$ to the end of the list $\vec v$.
\item
Replace $T$ with the set of elements of its elements which have a zero parity under the parity $p(x)$ just found: $T \rightarrow \left\{ x \mid p(x)=0 , x\in T\right\}$, and then go to Step 2.
\item
One could stop here!  The elements of  $\vec v$ are the components of a continuous ideal vector parity for objects in $S$.
\item
However, in an optional final `cleaning' step, one could look through the elements of $\vec v$ to see if any of the  parities added to it early on were made redundant by later additions.  The last parity added to $\vec v$ will never be removable in this way, but others added to the list earlier might be.
\end{enumerate}
Although Step 4 always removes elements from $T$, there is no \textit{guarantee} that this step alone will cause the algorithm to terminate since $S$ (and thus the $T$s) may contain infinitely many elements. However, when elements of $S$ have finitely many real degrees of freedom the algorithm is likely to terminate so long as the parities $p(x)$ are not chosen unreasonably.  The reason is that each parity $p(x)$ chosen in Step 3 must (because it is continuous) give a non-zero parity to not just a single element but to a continuous degree of freedom's-worth of elements $x\in T$. Consequently, each Step 4 has the capacity to reduce the dimension of T by around one.\footnote{This statements does not constitute a proof. It is at best only a plausibility argument; there are almost certainly many pathological ways one could try to choose continuous scalar parities $p(x)$ in Step 3 that would prevent the algorithm failing to terminate even when $S$ has finitely many real degrees of freedom. } 
In short, the message of  Lemma~\ref{lem:nouniversalparity}  is that in `reasonable' cases:\begin{quote}Sufficiently complex objects may require a \textit{vector} of scalar parities to describe their chirality if it is desired that:
(i) every chiral object shall have a non-zero parity (i.e.~the parity is ideal), and (ii) the parities are continuous functions on the space of objects.
\end{quote}
In more colloquial language:
\begin{quote}
The handedness of some sorts of objects are only describable only by a collection of numbers -- a single number is not always sufficient.
\end{quote}

\section{Concrete examples and discussions}
\label{sec:exampleseectreewre}

\subsection{Parities as chiral measures for origin-centred two dimensional star domains}
In this example we define `star domains' and explore what happens if one tries to use parities as measure of their chirality.  We contrast the effects of requiring the chiral measures to be  invariant with respect to rotations (or not) in Sections~\ref{examplenorotinvar} and \ref{examplerotinvar}.

\begin{figure}\begin{center}
%
%
  \includegraphics[width=0.4\linewidth]{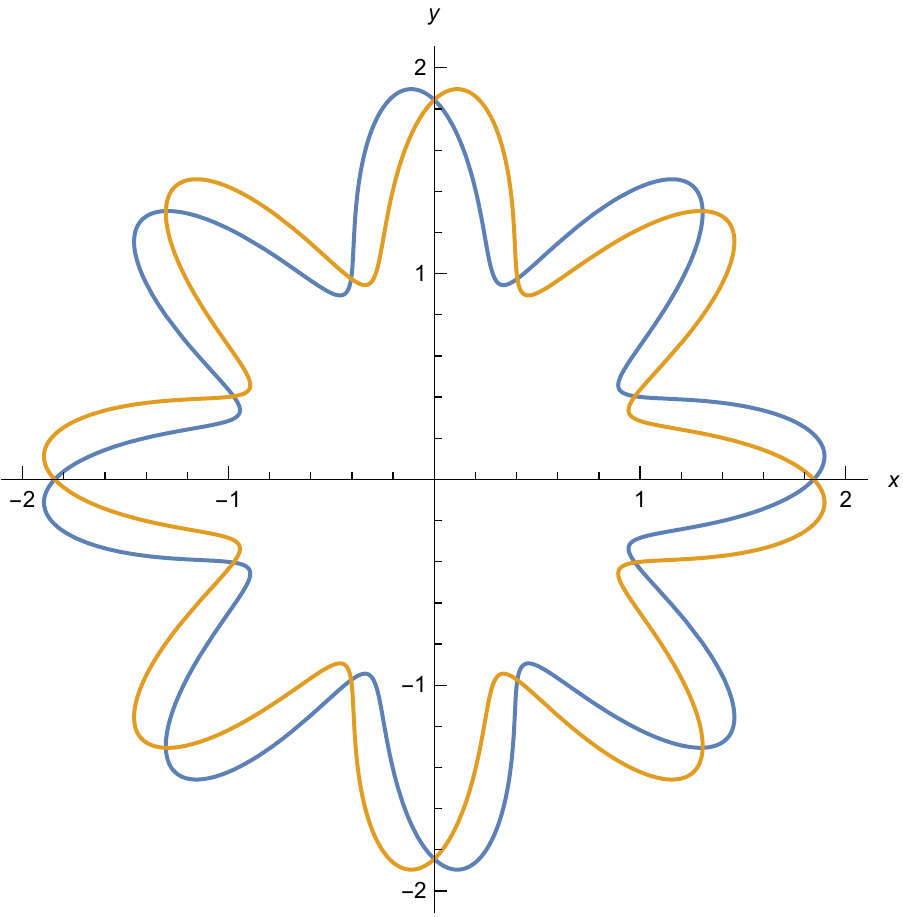}
    \caption{\label{fig:complexstar}
    This figure shows a pair of star-domains which are images of each other under the parity operator defined in Section~\ref{sec:wherestardomainparityisdefined}.  The shape hopefully makes apparent why the name `star-domain' is appropriate.
  }
  \end{center}
\end{figure}

Any continuous real function $r(\theta)$ which is always-positive and has period $2\pi$ can be interpreted as the polar representation of the boundary of an origin-centred two-dimensional star domain.\footnote{A set $S$ in  $\mathbb{R}^n$ is called a star domain if there exists a vantage-point $x$ in $S$ such that for all $y$ in $S$ the line segment from $x$ to $y$ is in $S$ \cite{wiki:stardomain}.
If the origin is a vantage-point for the set, we call $S$ an origin-centred star domain (OCSD).}
Examples of star domains may be found in Figures~\ref{fig:complexstar}, \ref{fig:simplestar} and \ref{fig:blobofdeath}.

Since any such function $r(\theta)$ can also be decomposed into Fourier Series, every star-domain can also be represented by a (possibly infinite) list of real coefficients $a_n$ and $b_n$ via:
\begin{align}
\label{eq:thingwithanbninitfirst}
r(\theta) &= a_0 + \sum_{n=1}^{\infty} a_n \cos{n\theta} + \sum_{n=1}^{\infty} b_n \sin{n\theta}.
\end{align}

\noindent [ Aside: The converse is not true: not all lists of coefficients describes a star domain. For example: $a_0=-1$ and all other coefficients zero does not describe the boundary of a star-domain since it leads to $r(\theta)<0$. ]

We will now calculate the action of parity and rotations of on the coefficients $a_n$ and $b_n$ needed to represent the star domains, and we will then use these coefficients to build parities which can act as measures of the chirality of these figures.

\subsubsection{The action of Parity on such figures}
If the action of parity on a \label{sec:wherestardomainparityisdefined} two-dimensional star domain is defined to be the reflection of the domain in the $x$-axis (i.e.~if parity has the effect of mapping $y\rightarrow -y$) then it is readily seen that this is the same as mapping $r(\theta)\rightarrow r(-\theta)$.  This is in turn the same as mapping $(a_n, b_n)\rightarrow (a_n, -b_n)$.  For such figures we therefore define the action of the parity operator, $P$, on the Fourier coefficients as follows:
\begin{align}
P \left(\begin{array}{c}
a_n \\ b_n
\end{array}\right) &= 
\left(\begin{array}{cc}
1 & 0 \\
0 & -1
\end{array}\right)
\left(\begin{array}{c}
a_n \\ b_n
\end{array}\right).
\label{eq:parityonexample2dfigs}
\end{align}
The $a_n$ are therefore even and the $b_n$ odd under parity.
\subsubsection{The action of Rotations (about the origin) on such figures}
When such a figure is rotated by an angle $\delta$, the effect:
\begin{align}
r(\theta) &\rightarrow r(\theta-\delta)
= 
a_0 + \sum_{n=1}^{\infty} a_n \cos{(n\theta - n\delta)} + \sum_{n=1}^{\infty} b_n \sin{(n\theta-n\delta)}
\nonumber
\\
&= 
a_0 + \sum_{n=1}^{\infty} a_n (\cos{n\delta}\cos{n\theta}+\sin{n\delta}\sin{n\theta}) + \sum_{n=1}^{\infty} b_n( \cos{n\delta}\sin{n\theta}-\sin{n\delta}\cos{n \theta})
\nonumber
\\
&=
a_0 + \sum_{n=1}^{\infty} [a_n \cos{n\delta}-b_n \sin{n\delta} ] \cos{n\theta}
+ 
\sum_{n=1}^{\infty} [a_n\sin{n\theta} + b_n \cos{n\delta}]  \sin{n\theta}
\nonumber
\end{align}
means that if $R_\delta$ is the operator for `rotation by angle $\delta$', then the effect of that rotation on the pair of coefficients $(a_n,b_n)$ is given by:
\begin{align}
\left(\begin{array}{c}
a_n \\ b_n
\end{array}\right)
\rightarrow
(R_\delta )^n \left(\begin{array}{c}
a_n \\ b_n
\end{array}\right) &= 
\left(\begin{array}{cc}
{\cos{n\delta}} & {-\sin{n \delta}} \\ 
{\sin{n\delta}} & {\cos{n \delta}}
\end{array}\right)
\left(\begin{array}{c}
a_n \\ b_n
\end{array}\right).\label{eq:parityonexample2dfigsrot}
\end{align}

\subsubsection{Continuous parities as measures of chirality for star-domains when orientation matters}
\label{examplenorotinvar}
Here we consider what happens when the symmetry group $S$ which defines the `boring' or `unimportant' transformations consists of just the identity: $S=\{e\}$.  This is equivalent to saying that `orientation matters' for the purposes of assessing whether objects are or are not the same as their images under parity. 

\begin{figure}\begin{center}
  \includegraphics[width=0.4\linewidth]{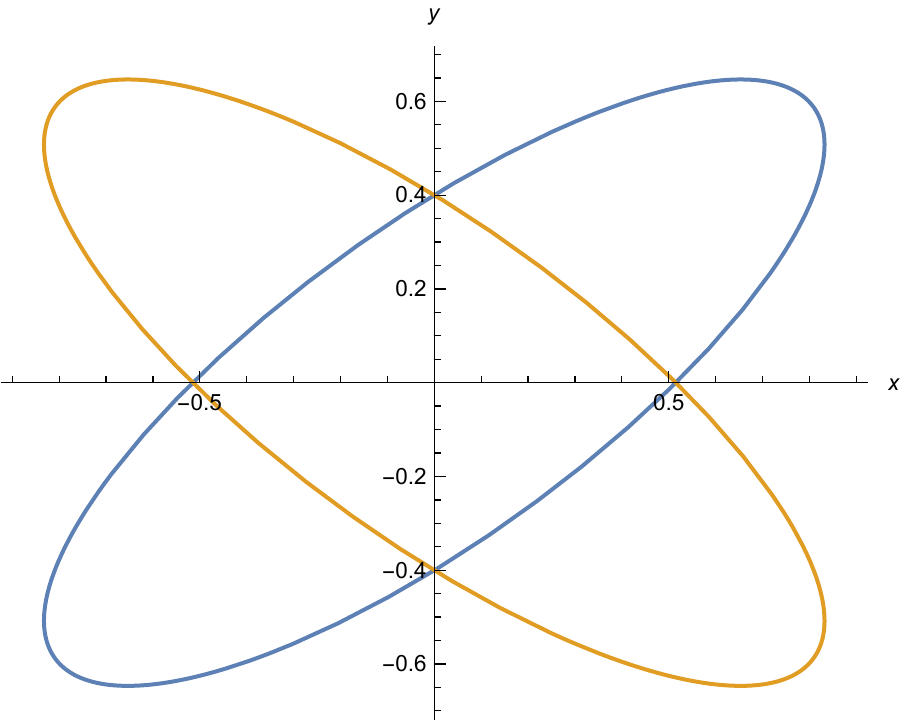}
    \caption{\label{fig:simplestar}
    A pair of ellipse-shaped star-domains which are images of each other under the parity operator defined in Section~\ref{sec:wherestardomainparityisdefined}.
  }
  \end{center}
\end{figure}

For example: when orientation matters the ellipse-shaped regions shown in Figure~\ref{fig:simplestar} are both considered chiral because each differs from its image under parity even though both ellipses would be identical if we could ignore their orientation.

Trivially, any of the values $b_n$ are odd under parity (as demonstrated in \eqref{eq:parityonexample2dfigs}) and so any of these $b_n$ would count as continuous parities which could be used as continuous chiral measures for origin-centred star domains when orientation matters.

\subsubsection{Continuous parities as measures of chirality for star-domains when orientation does \textbf{not} matter}
\label{examplerotinvar}
In contrast to Section~\ref{examplenorotinvar} we now consider what would happen when the symmetry group $S$ consists of all rotations about the origin.  In this case two objects are considered equivalent (for the purposes of determining their chirality) if they match their images under parity after rotation about the origin.  Or more plainly put, orientation can be ignored when comparing objects and their images under parity.

Though the Fourier coefficients $a_n$ and $b_n$ in \eqref{eq:thingwithanbninitfirst} are not individually invariant under rotations\footnote{This statement excludes the trivial cases of $a_0$ or any Fourier mode $i$ for which both $a_i=0$ and $b_i=0$.} it is possible to construct functions of the $a_n$ and $b_n$ which are rotationally invariant and yet remain odd under parity.
Indeed, an infinite number of such parity odd rotational invariants can be constructed.  Some examples (parameterised by positive integers $n$, $m$ and $t$) include:
\begin{align}
I_{n,m;t} &= 
\int_{\delta=0}^{2 \pi}
\int_{x=0}^{2 pi}
f^n(n,m,x) f^m(n,m,x+\delta) \sin{(t \delta)}
\ 
dx
\ 
d\delta
\label{eq:formulaforcorrelatorinvariant}
 \end{align}
 in which
 \begin{align}
f(n,m,x) &=
 b_n \sin{n x} + b_m \sin{m x} + a_n \cos{n x} + a_m \cos{m x}.
 \end{align}
 Up to constant factors, 
the low order rotationally invariant parities generated by the above formula include these:
\begin{align}
I_{1,2;1} &\propto
-2 a_1 a_2 b_1 + a_1^2 b_2 - b_1^2 b_2
\label{eq:firstexample}
,\\
I_{1,3;1} &\propto
   3 a_1 b_1^2 b_3
,\\
I_{2,3;1} &\propto
2 a_2^3 a_3 b_3 - 6 a_2 a_3 b_2^2 b_3 + 
   b_2^3 (a_3^2 - b_3^2) + 3 a_2^2 b_2 (-a_3^2 + b_3^2)
,\\
I_{3,4;1} &= 0,\\
I_{3,4;2} &\propto
4 a_3 a_4^3 b_3 (-a_3^2 + b_3^2) + 
   3 a_4^2 (a_3^4 - 6 a_3^2 b_3^2 + b_3^4) b_4 
   \nonumber \\ &\phantom{\qquad} 
   + 12 a_4 b_3 (a_3^3 - a_3 b_3^2) b_
     4^2 - (a_3^4 - 6 a_3^2 b_3^2 + b_3^4) b_4^3
     \label{eq:Ithreefourtwo}
,\\
I_{3,5;1} &\propto
-a_5^3 b_3 (5 a_3^4 - 10 a_3^2 b_3^2 + b_3^4) + 
   3 a_3 a_5^2 (a_3^4 - 10 a_3^2 b_3^2 + 5 b_3^4) b_5 
    \nonumber \\ &\phantom{\qquad} 
    + 3 a_5 b_3 (5 a_3^4 - 10 a_3^2 b_3^2 + b_3^4) b_5^2 - 
   a_3 (a_3^4 - 10 a_3^2 b_3^2 + 5 b_3^4) b_5^3
   .
\label{eq:lastexample}
\end{align}
The parity-oddness of the invariants \eqref{eq:firstexample} to \eqref{eq:lastexample} is  self evident from the odd number of powers of $b_n$ each term is seen to contain, and is enforced by the $\sin(t\delta)$ term in \eqref{eq:formulaforcorrelatorinvariant}. One could verify the rotational invariance of \eqref{eq:firstexample} to \eqref{eq:lastexample} by substituting the transformation \eqref{eq:parityonexample2dfigsrot} into each case separately, but it is much easier to see that  rotational invariance is a direct consequence  of the fact that  \eqref{eq:formulaforcorrelatorinvariant} can be viewed as a sum (over $\delta$) of $\delta$-dependent correlation functions, each computed from two functions of $x$ which share a common period of $2 \pi$.

\begin{figure}
\begin{center}
  \includegraphics[width=0.99\linewidth]{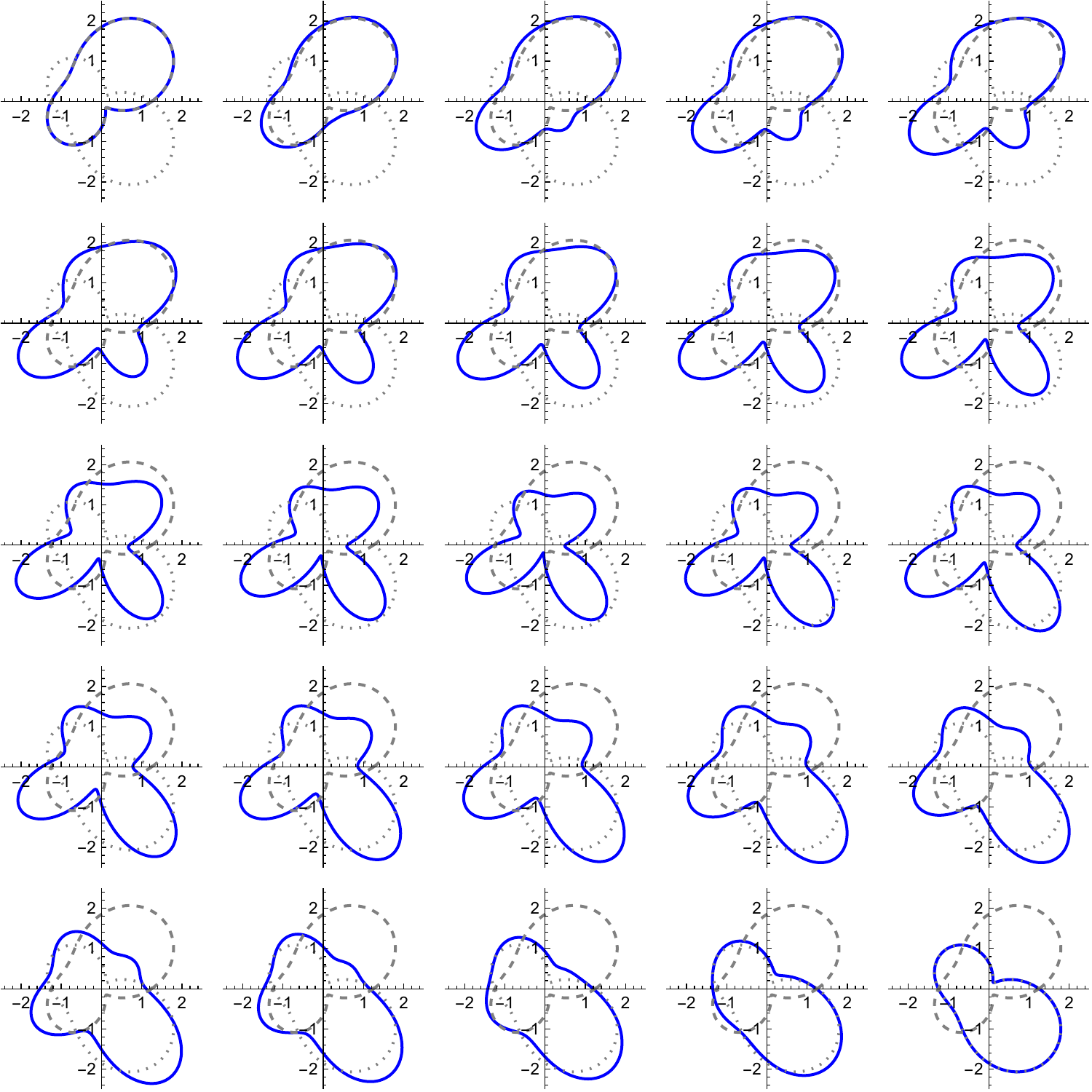}
    \caption{
    A example showing a continuous transformation of one star domain (top left, $r_{tl}(\theta)\equiv[25 + 12(\sin \theta +\sin {2\theta})]/20$) into to a mirror image of itself (bottom right, $r_{br}(\theta)\equiv r_{tl}(-\theta)$) via a chiral intermediate (centre,  ($r_m(\theta)=
    \frac 1 8 [10 - 3 \cos{3 x} - 3 \cos{4 x} - 3 \sin{3 x} + 3 \sin{4 x}
    ]
    $).  Between each `frame' a single parameter $t\in[0,1]$ (not labelled) advances by $\frac 1 {24}$ to interpolates between the three figures.  Concretely: $r(\theta;t)=(1-2t)\cdot r_{tl}(\theta) + 4 t(1-t)\cdot r_{m}(\theta)$ when $0\le t \le \frac 1 2$ and $r(\theta;t)=4 t(1-t)\cdot r_{m}(\theta) + (2t-1)\cdot r_{br}(\theta)$ when $\frac 1 2 \le t \le 1 $. Figure~\ref{fig:slideofdeath} proves that such a transformation does not pass through an intermediate achiral state, and so Figure~\ref{fig:blobofdeath} could supplant  Figure~\ref{fig:handofdeath} in its role in the proof of  Lemma~\ref{lem:nouniversalparity}.  In short: Figure~\ref{fig:blobofdeath} is a concrete realisation of the  transformation represented conceptually in Figure~\ref{fig:handofdeath}.\label{fig:blobofdeath} 
  }
  \end{center}
\end{figure}

\begin{figure}
\begin{center}
  \includegraphics[width=0.8\linewidth]{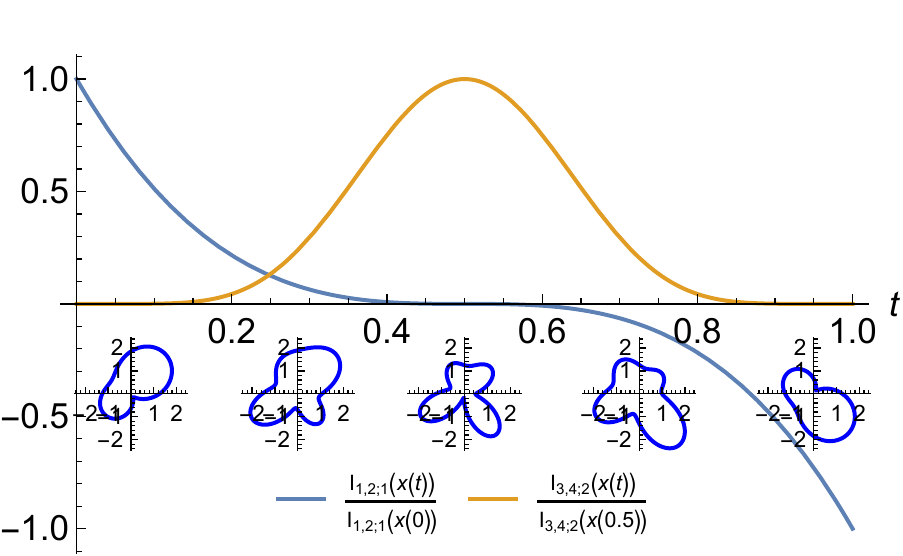}
    \caption{
    This plot shows how the rotationally invariant parities $I_{1,2;1}$ and $I_{2,4;2}$ (see \eqref{eq:firstexample} and \eqref{eq:Ithreefourtwo}) vary with $t$ as the star domain shown in the upper left corner of Figure~\ref{fig:blobofdeath} (represented by $t=0$) transforms into that shown in the lower-right of the Figure~\ref{fig:blobofdeath} ($t=1$). The interpolating states are described in the caption of  Figure~\ref{fig:blobofdeath} and examples for  $t\in\{0, 0.25, 0.5, 0.75, 1.0\}$ are inset for context.  The existence of at least one non-zero parity at all values of $t$ shows that this shape transformation does not pass through an intermediate achiral state. \label{fig:slideofdeath}
    Figure 7 of \cite{RevModPhys.71.1745} illustrates a similar process but in a  molecular chemistry context.
  }
  \end{center}
\end{figure}

Figure~\ref{fig:slideofdeath} shows how two of these parities vary as an example star-domain continuously transforms into its own mirror image via the route shown in Figure~\ref{fig:blobofdeath}.

\subsection{An example: collisions in particle physics}

The set 
$S_{ab\rightarrow jjj+X}$ consisting of ``particle-physics events with three jets (and some other stuff) in a final state resulting from the collision of an `$a$' with a `$b$'{}'' is not thought to admit a continuous ideal scalar parity.  Ref~\cite{Lester:2020jrg} was nonetheless able to use an algorithm similar to the one above to construct a 19-dimensional continuous ideal vector parity for  events in that set. 

For a fixed observer, events in  $S_{ab\rightarrow jjj+X}$ have 20 degrees of freedom as there are 5 four-momenta in total and each has 4 components.  However, three of degrees of freedom may be thought of specifying the observer velocity with respect to the collision, and another three fix the observer's orientation relative to it.  There are therefore 14 degrees of freedom \textit{intrinsic} to the particles themselves (rather than related to the observer's frame of reference).  The number of intrinsic degrees of freedom, 14,  is smaller than (but not far from) the number of scalar parities, 19, found in \cite{Lester:2020jrg}.   This suggests that the vector of parities found in that paper was not the most efficient set which could have been chosen -- but not far from it. This is consistent with \cite{Lester:2020jrg}'s claims that the set of parities is sufficient but not guaranteed to be the smallest possible.

\section{End Note}
It does not seem appropriate to include a conclusion or a summary section given that the paper does not advance an argument or thesis which would require either.  


\section*{Acknowledgements}
Special thanks are given to Rupert Tombs for assisting in the push to get this paper and two others out on a very short timescale.  Without the much appreciated encouragement of Lars Henkelman and other members of the Cambridge ATLAS group, it is unlikely that this work would have been made public following the discovery of \cite{RevModPhys.71.1745}.     The author acknowledges many forms of support from Peterhouse including (but not limited to) various forms of hospitality, very helpful discussions with Namu Kroupa and L\^e Nguy$\tilde{\hat{\mathrm{e}}}$n Nguy{\^e}n Kh\^oi, and unusual contributions from members of its MCR in the early morning of the eve of All Saints' Day. 
\printbibliography
\end{document}